\documentclass[11pt, a4paper, logo, copyright]{googledeepmind}

\usepackage[authoryear, sort&compress, round]{natbib}

\usepackage{amsmath,amsfonts,bm}

\def\eqref#1{equation~\ref{#1}}

\def\1{\bm{1}}

\DeclareMathAlphabet{\mathsfit}{\encodingdefault}{\sfdefault}{m}{sl}
\SetMathAlphabet{\mathsfit}{bold}{\encodingdefault}{\sfdefault}{bx}{n}

\usepackage{times}
\usepackage{latexsym}
\usepackage[T1]{fontenc}
\usepackage[utf8]{inputenc}
\usepackage{microtype}
\usepackage{inconsolata}
\usepackage{graphicx}
\usepackage{balance}       
\usepackage{graphics}      
\usepackage{hyperref}
\usepackage{color}
\usepackage{booktabs}
\usepackage{textcomp}
\usepackage{subcaption}
\usepackage{enumerate}
\usepackage{xcolor}
\usepackage{lipsum}
\usepackage{makecell}
\usepackage{multicol}
\usepackage{multirow}
\usepackage{array}
\usepackage{verbatimbox}
\usepackage{enumitem}
\usepackage{amsmath}
\usepackage{float}
\usepackage{stfloats}
\usepackage{graphicx}
\usepackage{amsthm}
\usepackage{listings}
\usepackage{caption} 
\usepackage[export]{adjustbox}
\usepackage{xspace}
\usepackage{epsfig}
\usepackage[linesnumbered]{algorithm2e}
\usepackage{algpseudocode}
\usepackage{tabularx}
\usepackage{arydshln}
\usepackage[bottom]{footmisc}
\usepackage{tcolorbox}
\usepackage{stackengine}
\usepackage{placeins}

\usepackage[normalem]{ulem}
\useunder{\uline}{\ul}{}
\tcbuselibrary{breakable}

\definecolor{E+F}{RGB}{	255, 99, 71}
\definecolor{B+F}{RGB}{255, 165, 0}
\definecolor{E+I}{RGB}{	173, 216, 230}
\definecolor{B+I}{RGB}{	30, 144, 255}
\definecolor{D}{RGB}{	60, 179, 113}

\definecolor{maroon}{cmyk}{0,0.87,0.68,0.32}

\usepackage{tikz}

\definecolor{darkgreen}{rgb}{0.0, 0.5, 0.0}
\definecolor{usercolor}{RGB}{200, 230, 250} 
\definecolor{coachcolor}{RGB}{180, 250, 180} 

\title{CGM Data Analysis 2.0: Functional Data Pattern Recognition and Artificial Intelligence Applications}
\reportnumber{} 
\renewcommand{\today}{2025-05-08}

\author[1,$\dagger$]{David C. Klonoff, MD, FACP, FRCP (Edin), Fellow AIMBE}
\author[2]{Richard M. Bergenstal, MD}
\author[3]{Eda Cengiz, MD, MHS, FAAP}
\author[4]{Mark A. Clements, MD, PhD}
\author[5]{Daniel Espes, MD, PhD}
\author[6]{Juan Espinoza, MD}
\author[7]{David Kerr, MBChB, DM, FRCP, FRCPE}
\author[8]{Boris Kovatchev, PhD}
\author[9]{David M. Maahs, MD, PhD }
\author[10]{Julia K. Mader, MD}
\author[11]{Nestoras Mathioudakis, MD, MHS}
\author[12,13]{Ahmed A. Metwally, PhD}
\author[14]{Shahid N. Shah, MSc}
\author[15]{Bin Sheng, PhD}
\author[16]{Michael P. Snyder, PhD}
\author[17]{Guillermo Umpierrez, MD}
\author[18]{Alessandra T. Ayers, BA}
\author[18]{Cindy N. Ho, BA}
\author[19]{Elizabeth Healey, PhD}

\affil[1]{Diabetes Research Institute, Mills-Peninsula Medical Center, San Mateo, CA, USA}
\affil[2]{International Diabetes Center at Park Nicollet, MN, USA}
\affil[3]{Department of Pediatrics, University of California, San Francisco, San Francisco, CA, USA}
\affil[4]{Children’s Mercy Kansas City, Kansas City, MO, USA}
\affil[5]{Department of Medical Cell Biology, Department of Medical Sciences, and Science for Life Laboratory, Uppsala University, Sweden}
\affil[6]{Stanley Manne Children’s Research Institute, Ann \& Robert H. Lurie Children’s Hospital of Chicago, Chicago, IL, USA}
\affil[7]{Center for Health Systems Research, Sutter Health, Santa Barbara, CA, USA}
\affil[8]{Center for Diabetes Technology, School of Medicine, University of Virginia, Charlottesville, VA, USA}
\affil[9]{Department of Pediatrics, Stanford University, Stanford, CA, USA}
\affil[10]{Division of Endocrinology and Diabetology, Department of Internal Medicine, Medical University of Graz, Graz, Austria}
\affil[11]{School of Medicine, Johns Hopkins University, Baltimore, MD, USA}
\affil[12]{Google Research, Mountain View, CA, USA}
\affil[13]{Systems and Biomedical Engineering Department, Cairo University, Giza, Egypt}
\affil[14]{Netspective Foundation, Inc., Silver Spring, MD, USA}
\affil[15]{Department of Computer Science and Engineering, Shanghai Jiao Tong University, Shanghai, China}
\affil[16]{Department of Genetics, Stanford University, Stanford, CA, USA}
\affil[17]{Division of Endocrinology, Department of Medicine, Emory University School of Medicine, Atlanta, GA, USA}
\affil[18]{Diabetes Technology Society, Burlingame, CA, USA}
\affil[19]{Boston Children’s Hospital, Harvard Medical School, Boston, MA, USA}

\affil[$\dagger$]{Correspondence: dklonoff@diabetestechnology.org}

\begin{abstract}

New methods of CGM data analysis are emerging that are valuable for interpreting CGM patterns and underlying metabolic physiology. These new methods use functional data analysis and artificial intelligence (AI), including machine learning (ML). Compared to traditional metrics for evaluating CGM tracing results (CGM Data Analysis 1.0), these new methods, which we refer to as CGM Data Analysis 2.0, can provide a more detailed understanding of glucose fluctuations and trends and enable more personalized and effective diabetes management strategies once translated into practical clinical solutions.  

\end{abstract}
\begin{document}

\maketitle
\section{Introduction}

Continuous glucose monitoring (CGM) has transformed diabetes management and is now often considered a standard of care for insulin treated diabetes. Recently, interest in CGM has extended into populations with and at-risk of non-insulin treated diabetes, and in 2024, three over-the-counter CGM devices were approved in the United States ~\citep{shah2025normal}. As these devices become more ubiquitous, a major challenge will be to make sense of the 1440 interstitial glucose readings that can be collected daily.

CGM data interpretation currently employs four main methodologies, each with distinct approaches and applications. These four analytical frameworks include: 1. traditional summary statistics, 2. Functional Data Analysis, 3. artificial intelligence (AI) / machine learning (ML), and 4. foundation models in AI. The first two of these methods use statistics to identify patterns of glycemia and are known as pattern analysis methods; the latter two methods involve AI. While traditional summary statistics simplify data into aggregate metrics, Functional Data Analysis uses advanced statistical methods to analyze temporal dynamics that can reveal more detailed physiological patterns.

Until recently, traditional statistical methods have been the predominant approach to analyzing CGM data and can be thought of as "CGM Data Analysis 1.0."  More recently, new Functional Data Analysis methods have been applied to identity patterns~\citep{gecili2021functional}, and AI models~\citep{medanki2024artificial}, which include ML methods~\citep{jacobs2023artificial}, have been developed for risk stratification. To identify trends, excursions, and variability in glucose levels, these new methods have the potential to provide additional insights beyond traditional pattern analyses of the three panels of an ambulatory glucose profile (AGP), which presents 1) summary glucose statistics and targets, 2) the ambulatory profile, and 3) daily glucose profiles~\citep{gecili2021functional}. Functional Data Analysis for pattern analysis and emerging AI-based and ML-based interpretation methods can be considered as "CGM Data Analysis 2.0". The advantage for clinicians involved in diabetes care will be access to more nuanced patterns of glycemia, which are foundational for personalizing diabetes management. This article discusses the features of CGM Data Analysis 1.0 and 2.0 and explains the progression from emphasis on traditional statistics to Functional Data Analysis to AI- and ML-based methods for interpreting CGM data patterns.

\section{WHY WE ARE MOVING BEYOND TRADITIONAL STATISTICAL METHODS}

Traditional pattern analysis of the CGM signal focuses on simple-to-calculate summary characteristics over 10-14 days for assessing glycemia or quality of glycemic control and is used widely by clinicians.  These metrics include the percentages of time spent in five glycemic ranges, the Glucose Management Indicator (which is proportional to the mean glucose concentration), and the coefficient of variation (which is a measurement of glycemic variability). These metrics tend to oversimplify dynamic glucose fluctuations and lack granularity in capturing complex temporal patterns. In contrast, Functional Data Analysis~\citep{gecili2021functional}, AI~\citep{medanki2024artificial}, and ML~\citep{kapoor2025continuous} are complex frameworks that use the entire CGM time series. Functional Data Analysis, compared to traditional statistical analysis, provides additional insights into the temporal structure of glycemic variability and offers greater emphasis on deconstructing the amount and timing of recurring variations during the entire wear period~\citep{gecili2021functional}. As such, Functional Data Analysis goes beyond traditional statistics to 1) present a more comprehensive analysis of glucose data, whereby complex metrics may supplement traditional glucose metrics, 2) allow sophisticated time-dependent observations (such as different patterns on weekdays vs. weekends), and 3) enable identification of phenotypes or subphenotypes with distinct postprandial or nocturnal glycemic patterns. ML and AI are able to both analyze complex glucose patterns and combine the analysis with personalized decision-making frameworks. ML algorithms have been used to analyze CGM data patterns to predict metabolic subphenotypes~\citep{metwally2024prediction} and predict future glycemic trends, whereas additional AI analyses can integrate these predictions with other health parameters to automate therapeutic interventions, such as closed loop control~\citep{eghbali2024application, mittal2025harnessing}. Although no AI-powered AID system is currently on the market, such a system has been successfully tested~\citep{kovatchev2024neural}. This approach to CGM data analysis also allows the algorithm to learn from the person living with diabetes and will reduce computational demands~\citep{Swensen2024AddingAI}. Table \ref{tab:table_1} compares key features for pattern analysis methods of CGM data using traditional statistical methods, Functional data Analysis, ML, and AI.

\begin{table}[htbp]
\centering
\caption{A comparison of key features for pattern analysis methods of CGM data using traditional statistical methods, Functional Data Analysis, ML, and AI. Abbreviations: AI, artificial intelligence; CGM, continuous glucose monitor; EHR, electronic health record; ML, machine learning.}
\label{tab:table_1}
\footnotesize
\begin{tabular}{ >{\raggedright\arraybackslash}p{0.12\textwidth} 
                 >{\raggedright\arraybackslash}p{0.20\textwidth} 
                 >{\raggedright\arraybackslash}p{0.20\textwidth} 
                 >{\raggedright\arraybackslash}p{0.20\textwidth} 
                 >{\raggedright\arraybackslash}p{0.20\textwidth}  
                 }
\hline 
\textbf{Method}           & Traditional Statistical Pattern Analysis & Functional Data Pattern Analysis                                                      & Machine Learning Pattern Analysis                                               & Artificial Intelligence Pattern Analysis                                              \\ \hline 
\textbf{Reference}        & Scheiner et al.~\citep{scheiner2016cgm}                                 & Gecili et al.~\citep{gecili2021functional}                                                                                 & Jacobs et al. and Metwally et al.~\citep{jacobs2023artificial, metwally2024prediction}                                                     & Shomali et al.~\citep{shomali2024development}                                                                               \\ \hline 
\textbf{Approach}         & Visual, summary statistics                        & Statistical, models entire time series                                                         & Predictive modeling using algorithms and glucose timeseries                              & Integrates machine learning, deep learning, and advanced algorithms                            \\ \hline 
\textbf{Data Used}        & Aggregated, summary, or graphical                 & Each CGM trajectory is considered a random function                                            & Large CGM datasets                                                                       & Massive, heterogeneous datasets (CGM, EHR, images, lifestyle, genomics)                        \\ \hline 
\textbf{Purpose}          & Identify obvious trends/patterns                  & Quantify, compare, and model complex dynamics                                                  & Predict future glucose levels and classify states (e.g., metabolic subphenotypes)        & Predict risk, classify subtypes, optimize therapy                                             \\ \hline 
\textbf{Main Users}       & Clinicians (practical use)                        & Statisticians, researchers                                                                     & Data scientists, AI/ML digital health researchers                                        & Researchers, health systems, digital therapeutics developers                                   \\ \hline 
\textbf{Depth of Insight} & Moderate (can identify trends and outliers)       & High (treats glucose trajectories as mathematical functions rather than discrete measurements) & High (can uncover non-linear, complex, hidden patterns)                                  & Very high (enables real-time adaptive interventions)                                           \\ \hline 
\textbf{Examples}         & AGP, time-in-range, mean, SD, GMI, GRI            & Functional principal components, glucodensity                                                  & Clinically meaningful patterns from complex CGM data,                                    & AI-powered CGM or AI-powered closed-loop insulin delivery, image-based complication detection \\ \hline 
\textbf{Limitations}      & May miss subtle/intricate patterns                & Requires statistical expertise, more complex                                                   & Requires large dataset to avoid overfitting and not generalizing beyond the used dataset & Data privacy, bias, transparency, regulatory hurdles, and a need for extensive validation      \\ \hline 
\end{tabular}
\end{table}

\section{TRADITIONAL STATISTICAL METHODS FOR CGM PATTERN ANALYSIS}

Traditional statistical methods for CGM pattern analysis focus on summarizing individual glycemic profiles, assessing variability, and identifying clinical events. These approaches prioritize aggregated metrics and risk indices, but do not provide insights into temporal trends. Traditional statistics focus on summary metrics such as mean glycemia, percentage of time in various glycemic ranges, and the amount of variability of the entire series. Seven traditional statistics are presented in the Ambulatory Glucose Profile~\citep{bergenstal2013recommendations}. Several composite metrics have been derived from these traditional statistics, including the Glycemia Risk Index~\citep{klonoff2023glycemia}, the Low/High Blood Glucose Indices (LBGI/HBGI)~\citep{kovatchev2002methods, kovatchev1998assessment}, and the glucose pentagon~\citep{nguyen2020review}. Overall, traditional statistics emphasize summaries and risk scores that are easily understood by clinicians but do not account for short-term glucose excursions and dynamic patterns, which could provide a more detailed picture of glycemic variability than traditional summary~\citep{matabuena2023reproducibility}. Also, these traditional summary statistics for pattern analysis are prone to distortion from missing data or irregularly spaced measurements due to sensor or connectivity problems and can fail to capture nuanced phenotypes. A review of ambulatory profiles and daily glucose profiles may sometimes identify day and night differences and times when there is glucose variability, but many clinicians are too time-limited to perform this type of pattern review. Traditional analysis of dense time-series CGM data can oversimplify patterns and has been increasingly supplemented with or supplanted by advanced statistical techniques known as Functional Data Analysis~\citep{gecili2021functional}. These complex statistics, compared to traditional statistics, can more accurately classify nuanced patterns, account for glucose dynamics over time, identify phenotypes, and facilitate personalization.

\section{WHEN TRADITIONAL STATISTICAL METHODS MAY NOT BE SUFFICIENT FOR PATTERN ANALYSIS}

Three physiological rationales support the use of Functional Data Analysis, ML, and AI, compared to traditional statistical methods, to provide more granular and mechanistic insights for analysis of CGM patterns. First, the shape of the glucose curve reflects underlying pathophysiology. Glucose dynamics, especially postprandial glucose responses, depend on numerous physiological parameters, including insulin sensitivity and beta cell function. Therefore, although two daily glucose curves at first glance can have a similar appearance, their differences may represent very different underlying pathophysiology, especially when considering also the use of CGM in pre-diabetes and type 2 diabetes (T2D)~\citep{barua2022northeast}. Functional Data Analysis outperforms traditional methods in capturing glucose patterns by modeling CGM trajectories as dynamic processes rather than static summaries~\citep{cui2023investigating}. Second, the shape of the glucose curve also can reflect the patient’s behaviors and medication regimens. For example, pattern analysis of postprandial glucose curves could reveal times when patients exercise, miss boluses of insulin, mistime boluses, alter insulin sensitivity by exercising, or incorrectly identify the macronutrient content of their food choices~\citep{cobry2010timing}. The Abbott Libreview app contains an example of data pattern analysis by including a section called "Mealtime Patterns" that displays overall trends in glucose levels before and after meals, broken down by time of day: morning, midday, evening, and night~\citep{BeyondType1LibreView}. However, this type of display does not meet the definition of Functional Data Analysis because it does not employ advanced statistical modeling of entire data trajectories as functions~\citep{gecili2021functional}. Understanding patterns and amounts of glycemic variability, risk of hypoglycemia and hyperglycemia, and factors contributing to variability allows identification of periods of increased risk to facilitate proactive management. Third, recognizing specific patterns in glucose levels, rather than the standard overall percentage of time within the target range or even a composite metric, can allow clinicians to identify the underlying root cause and help them to adjust treatments based on identifying times of day when glucose levels are likely to be out of personal target ranges~\citep{choudhary2013blood}.

\section{FUNCTIONAL DATA ANALYSIS FOR CGM PATTERN RECOGNITION}
Functional Data Analysis leverages the full time-series structure, and is much more powerful than traditional statistical pattern analysis, especially when the goal is understanding the full temporal dynamics of glucose fluctuations rather than relying on isolated summary statistics or discrete time points. This approach treats CGM data as dynamic curves rather than discrete points, allowing for insights into glucose patterns~\citep{gecili2021functional}. Five indications for Functional Data Analysis include: 1) recognizing longitudinal or repeated measures for when CGM data is collected over multiple days or weeks to analyze the patterns and variability both within and between individuals over time, 2) phenotyping and subgroup identification to identify distinct glycemic phenotypes or subgroups based on the shape and variability of glucose curves, which can help tailor interventions for patients at higher risk for complications, 3) assessing the impact of meals or interventions by reviewing the entire postprandial glucose trajectory, rather than just a single outcome, 4) qualifying inter- and intra-day reproducibility of glucose patterns to test a CGM device’s precision, or 5) assessing glycemic variability of an individual as a risk factor for complications as part of a precision medicine approach. Thus, Functional Data Analysis compared to traditional statistics is better suited for capturing dynamic glucose patterns, handling complex data structures, and making time-dependent predictions. Functional Data Analysis improves CGM prediction accuracy by transforming raw glucose traces into functional objects that preserve temporal dependencies, dynamic trends, and individual heterogeneity. 

\begin{figure}[htbp] 
    \centering
    \includegraphics[width=0.88\textwidth]{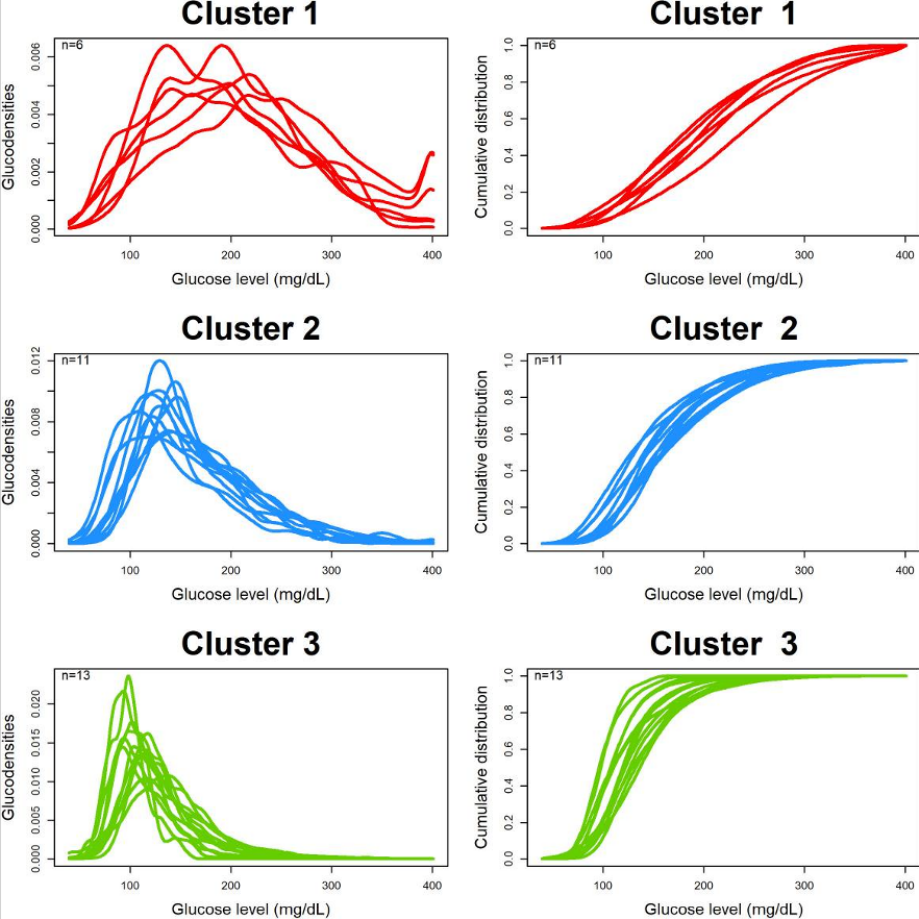}
    \caption{Estimated and clustered glucodensities (left three panels) and the corresponding cumulative distribution functions (right three panels). Three clusters are identified: red (six subjects), with the highest average and most variable levels of blood glucose; blue (11 subjects), with the somewhat better glycemic control; and, green (13 subjects), with the lowest average and least variable levels of blood glucose. Reproduced from Cui et al~\citep{cui2023investigating}. under the CC-BY-4.0 license (https://creativecommons.org/licenses/by/4.0/).}
    \label{fig:figure1}
\end{figure}

An example of Functional Data Analysis is the calculation of Glucodensity, a statistical approach for pattern analysis of CGM data, where the entire distribution of glucose values over time for an individual is represented as a probability density function~\citep{cui2023investigating, matabuena2021glucodensities}. Rather than focusing on summary statistics (such as mean glucose or time in range), with glucodensity the range intervals simultaneously shrink in width so that the new profile measures the proportion of time each patient spends at each specific glucose concentration rather than the amount of time spent within a wide range of glycemia. This method characterizes the full spectrum and variability of glucose concentrations, capturing both central tendency and fluctuations throughout the monitoring period~\citep{matabuena2024glucodensity}. A set of data analyzed with glucodensities is presented in Figure \ref{fig:figure1}.

A fundamentally different type of ML pattern recognition was used by Kovatchev and colleagues to add virtual CGM data to the Diabetes Control and Complications Trial (DCCT) results~\citep{kovatchev2025virtual}. In this study, the patterns of all 1,400 participants in the DCCT over ten years using their episodic Hemoglobin A1c readings and capillary glucose profiles were used to fill the gaps with CGM data derived from the previously identified CGM motifs~\citep{kovatchev2023clinically}. This study is an example of glucose pattern recognition for a different purpose – to upsample the data density. An ML-based method for defining six distinct CGM fluctuation patterns of glycemic variability and the durations in these patterns was reported by Chan and colleagues. These patterns described variability better than traditional statistical methods~\citep{chan2023time}.

\begin{figure}[htbp]
    \centering
    \includegraphics[width=0.88\textwidth]{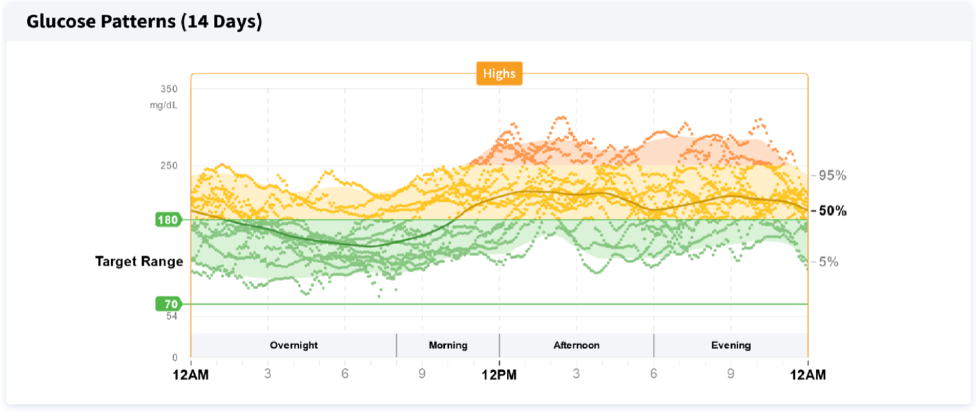}
    \begin{minipage}{\textwidth}
        \caption{A modal day plot (also known as a 14-day glucose pattern report) of a set of CGM tracings. The dark line is the mean glucose in any given time. (Figure is courtesy of Amiad Fredman.)}
        \label{fig:figure2}
    \end{minipage}
\end{figure}

\begin{figure}[htbp]
    \centering
    \includegraphics[width=0.88\textwidth]{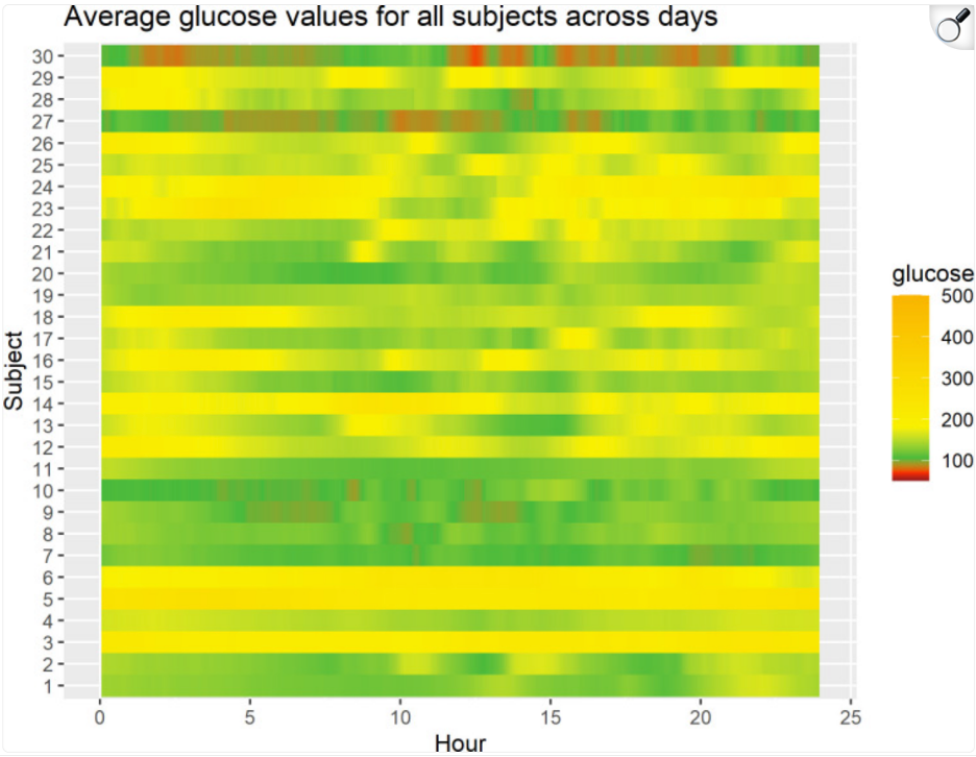}
    \begin{minipage}{\textwidth}
        \caption{A heat map of a set of CGM tracings from 30 subjects. For any given subject, at any given time point in the 0- to 24-h range, the average glucose level over study days corresponding to that time point for that subject is displayed. Reproduced from Cui et al.~\citep{cui2023investigating} under the CC-BY-4.0 license (https://creativecommons.org/licenses/by/4.0/).}
        \label{fig:figure3}
    \end{minipage}
\end{figure}

A modal day plot (also known as a 14-day glucose pattern report) is a visualization tool used with Functional Data Analysis, for longitudinal or repeated-measures data~\citep{desouter2025repeated}. Figure \ref{fig:figure2} presents an example of this type of plot. Although this type of plot is not a traditional statistical summary, it is often used in conjunction both with traditional statistics to inform statistical modeling and for Functional Data Analysis to visualize functional curve-based data~\citep{swihart2010lasagna}. A heat map can be used to present multiple stacked subjects’ CGM data across time using color gradients~\citep{swihart2010lasagna}. This plot displays the average hourly glucose concentration over the study days corresponding to that hour for that subject. With a heat map, one can observe differences, both between subjects and within subjects, as indicated by different colors in the heatmap. A heat map is presented in Figure \ref{fig:figure3}. This type of plot, however, does not demonstrate improving or worsening glycemic trends over time for the individual, because the plots demonstrate only glycemic averages for a given time of day and are considered a bridging tool between traditional statistics and Functional Data Analysis techniques~\citep{piersanti2023software}.

\section{ARTIFICIAL INTELLIGENCE AND MACHINE LEARNING FOR CGM PATTERN ANALYSIS}

There has been recent excitement about applying AI to analyze and interpret CGM data. AI encompasses a breadth of methodologies, from ML models that can learn characteristics and relationships in data to autonomous, generative systems that can independently analyze data. Several ML architectures, including recurrent neural networks, convolutional neural networks, and transformers, are capable of learning temporal patterns in time series data, similar to Functional Data Analysis.

However, AI offers additional capabilities as compared to Functional Data Analysis. For example, AI models can predict clinical outcomes and associate CGM data with clinical characteristics for tasks including risk stratification, subgroup identification, and decision support~\citep{medanki2024artificial}. ML models can use either raw CGM data or leverage existing pattern recognition techniques, including Functional Data Analysis, to gain clinical insights such as identifying clinically significant events (including impending hypoglycemia), assigning severity scores, and linking CGM characteristics to clinical phenotypes~\citep{giammarino2024machine, van2025feasibility}. 

AI also encompasses autonomous, intelligent systems that can be used to enhance the interaction between professionals and people with diabetes using CGM data. This includes interfaces that incorporate narrative summaries of data and intelligent insights personalized to an individuals' data. 

AI models can also employ predictive algorithms for real-time decision support. These models are used for short-term predictions, as part of closed loop systems, and for risk stratification. ML is primarily used to learn from CGM data to predict or classify glucose patterns, with a focus on enhancing risk stratification, subtype identification, root cause analysis, and prediction of patterns indicating adverse glycemic events~\citep{cichosz2024explainable}. The most advanced CGM interpretation systems today leverage both ML for prediction and AI for explanation, automation, and user interaction. As these two computational methods become combined, there will be the potential for optimizing diabetes management.

While predictive performance continues to improve, future clinical acceptance of AI-enhanced CGM tools will, of course, depend on their explainability and auditability. Clinicians need to be able to understand why an algorithm flagged a pattern or made a recommendation, especially in safety-critical scenarios such as insulin dosing. Explainable AI methods, such as attention mapping in deep learning models or SHAP values in ensemble approaches, can support transparency and trust in clinical decision-making. However, it will take significant experimentation to arrive at the best techniques for verification and validation, so moving from “ideation” to active trials and implementation in real-world scenarios is crucial.

Another emerging challenge is that of model drift and decay—when changes in physiology, lifestyle, or medication regimens reduce model accuracy over time. Techniques such as automated drift detection, performance monitoring, and continuous learning pipelines will be needed to maintain robustness in real-world deployments. Additionally, AI tools must account for edge-case populations who may fall outside the dominant training distributions, such as those with disrupted circadian rhythms, polypharmacy, or comorbidities. Like explainability and auditing, model drift and decay detection will take substantial experimentation. Putting AI into real-world use is crucial so that reinforcement learning with human feedback (RLHF) can get started. Federated personalization and meta-learning approaches may allow models to rapidly adapt to these complex cases with minimal additional data.

Below we highlight examples of the use of AI and ML to analyze glycemic patterns, including: 1) Pattern Recognition and Event Classification Models, 2) Creation of Glucotypes, 3) ML to predict metabolic subphenotypes, 4) Large Language Models (LLMs) for CGM Data Summarization, 5) Examples of Commercial AI-Enhanced CGM Systems, 6) CGM foundation models, and 7) ML models to predict clinical outcomes from CGM.

\section{PATTERN RECOGNITION AND EVENT CLASSIFICATION MODELS}
Pattern recognition and event classification models using an automated AI-driven system specifically designed to detect and classify clinically significant CGM patterns (CGM events) use algorithms to identify these events based on signal shape, temporal features, and glucose categories at the start and end of each event. Such a system has been validated against expert clinician assessments and demonstrated high accuracy in event detection and classification~\citep{shomali2024development}. Machine learning for assessing glycemic status and risk prediction has been used with random forest and support vector machine models to predict nocturnal hypoglycemia~\citep{afentakis2025development}. Long short-term memory (LSTM) networks and convolutional neural networks (CNNs) have also been applied to CGM time-series data for hypoglycemia prediction by leveraging the temporal dynamics of glucose fluctuations to accurately predict adverse events, guide clinical interventions~\cite{shao2024generalization}, and identify underlying root causes~\citep{cederblad2023classification}.

\begin{figure}[ht]
    \centering
    \includegraphics[width=0.88\textwidth]{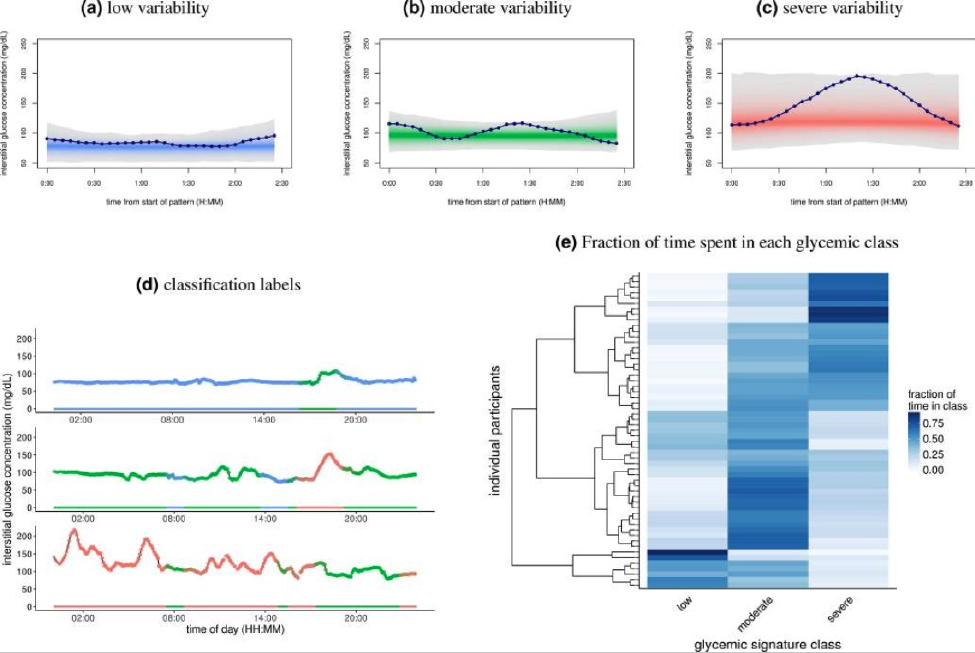}
        \caption{(a-c) Segregation of the 2.5-hour windows into the three classes of glycemic signatures derived from spectral clustering. The lines in each panel show an example of the glycemic signatures in each class. This separation of windows explains approximately 73\% of the variance. (d) One day of CGM data for 3 separate individuals. Color indicates classification of glycemic signatures. Note that since overlapping windows were used for clustering and classification, some periods of the day have multiple classifications. (e) Heat map showing the fraction of time individuals spent in each of the glycemic classes. Rows represent unique individuals in the cohort, while columns represent each of the glycemic signature classes shown in a-c. Color of the tiles corresponds to the fraction of time spent in each class, with 1 being 100\% of the time. There were 238 windows per participant (S3 Data). Rows of individuals are arranged according to hierarchical clustering. Abbreviations: CGM, continuous glucose monitoring. Reproduced from Hall et al~\citep{hall2018glucotypes}. under the CC-BY-4.0 license (https://creativecommons.org/licenses/by/4.0/).}
        \label{fig:figure4}
\end{figure}

\section{GLUCOTYPES}
AI algorithms using pattern analysis emphasizing CGM data glycemic variability have been used to identify subtypes of prediabetes and T2D. By defining three patterns of glycemic responses to standardized meals in people without known T2D, healthy people with no history of diabetes and normal static tests of glycemia (such as fasting plasma glucose, 2-hour plasma glucose following an oral glucose tolerance test [OGTT], or hemoglobin A1c concentration) can be categorized into one of three patterns of glucose metabolism to create “glucotypes”. Individuals with aberrant glucose metabolism, including even true T2D, can be identified with this approach~\citep{hall2018glucotypes, metwally2024predicting, metwally2024prediction}. Other investigators have worked on identifying glucotypes of people with diabetes by clustering patients’ CGM data. Investigators have described CGM tracings by the subgroup with which they best fall and have shown that this approach can delineate individuals with distinct statistical features and phenotypes~\citep{kovatchev2023clinically, tao2021multilevel, shao2024clinically, quan2022stratification}. A classification process for three glucotypes is presented in Figure \ref{fig:figure4}.

\section{ML-BASED ANALYSIS OF GLUCOSE TIMESERIES FOR PREDICTING METABOLIC SUBPHENTOYPES}

\begin{figure}[htbp]
    \centering
    \includegraphics[width=0.88\textwidth]{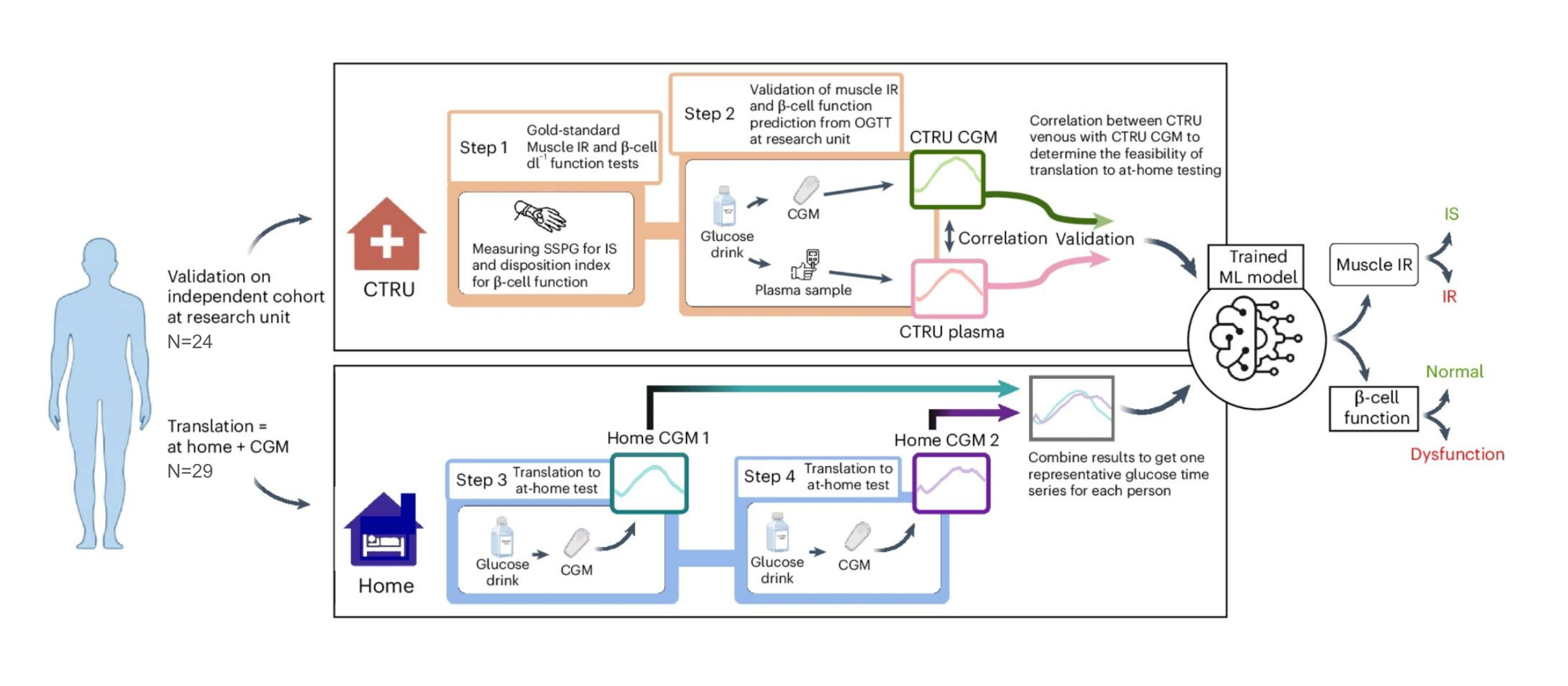}
    \begin{minipage}{\textwidth}
        \caption{Study design of the validation cohort and at-home OGTT test via CGM to predict muscle IR and $\beta$-cell function. Participants underwent gold-standard testing at the research unit for insulin resistance (SSPG test) and B-cell function (16-point OGTT with C-peptide deconvolution adjusted for SSPG and expressed as DI), as well as two OGTTs administered at home under standardized conditions during which glucose patterns were captured by a CGM within a single 10-day session (Dexcom G6 pro). Abbreviations: CGM, continuous glucose monitor; CTRU, clinical translational research unit; DI, Disposition Index; IR, insulin resistance; IS, insulin sensitivity; ML, machine learning; N, number of subjects; SSPG, steady-state plasma glucose. Reproduced from Metwally et al.~\citep{metwally2024prediction} under the CC-BY-4.0 license (http://creativecommons.org/licenses/by/4.0/).}
        \label{fig:figure5}
    \end{minipage}
\end{figure}

ML-based analysis of glucose time series for predicting metabolic subphenotypes has been used to directly predict metabolic subphenotypes, such as insulin resistance and beta cell function~\citep{metwally2024prediction}. Characterizing insulin resistance and beta cell function is of great interest, as it could allow for more targeted treatments. The gold standard test for insulin resistance is the hyperinsulinemic euglycemic clamp~\citep{defronzo1979glucose}, and for beta-cell function is the disposition index~\citep{bergman1981physiologic}. Both of these tests are performed in research facilities only, are expensive, and are time-consuming. In recent work, ML models have been trained on glucose time series from CGM following at-home OGTT to predict muscle insulin resistance and beta-cell function, which were measured using gold standard tests~\citep{metwally2024prediction}. This process is illustrated in Figure \ref{fig:figure5}. At-home identification of metabolic subphenotypes using a CGM therefore has the potential to facilitate risk stratification of individuals with early glucose dysregulation. It was shown that insulin resistance can be predicted at home with CGM and standardized meals~\citep{wu2025individual, park2024lifestyle}, or via wearables and routine blood biomarkers~\citep{metwally2025insulinresistancepredictionwearables}.

\begin{figure}[htbp]
    \centering
    \includegraphics[width=0.88\textwidth]{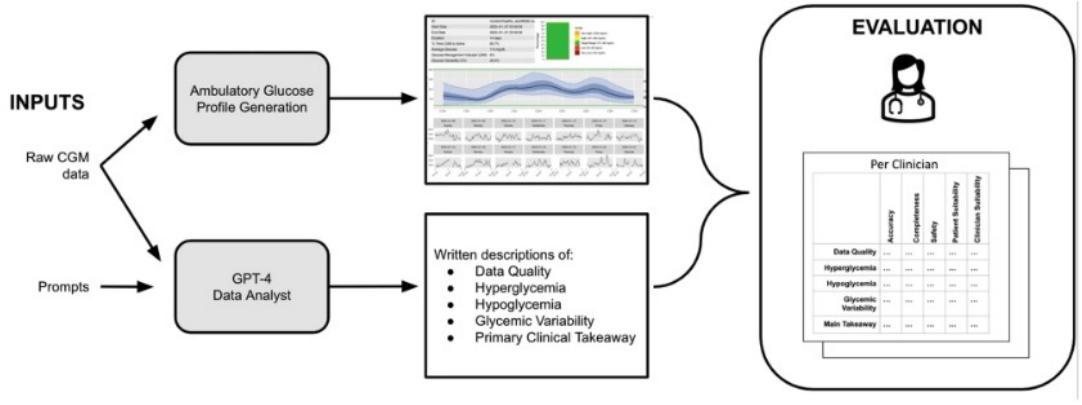}
    \begin{minipage}{\textwidth}
        \caption{Evaluation procedure for a single case of using GPT-4 to summarize an AGP. Figure reproduced from Healey et al.~\citep{healey2025case} under the CC-BY-4.0 license (http://creativecommons.org/licenses/by/4.0/).}
        \label{fig:figure6}
    \end{minipage}
\end{figure}

\section{LARGE LANGUAGE MODELS FOR CGM DATA SUMMARIZATION}
Large language models can analyze raw CGM data and generate narrative summaries similar to clinician-written reports~\citep{ho2024qualitative}. Healey and colleagues evaluated the ability of GPT-4 to compute quantitative metrics specific to diabetes found in an AGP as defined by an international consensus report~\citep{battelino2019clinical} and an American Diabetes Association Standards of Care in Diabetes report~\citep{care2023standards}. Qualitative summaries of the data in their AGP reports were derived from an article on interpreting an AGP report~\citep{czupryniak2022ambulatory}. They evaluated the accuracy, completeness, safety, and suitability of qualitative descriptions produced by GPT-4, as assessed by two clinician graders. GPT-4 provided qualitative descriptions of glycemic patterns, hypoglycemia, and hyperglycemia events. An evaluation procedure for a single case of using GPT-4 to summarize an AGP is presented in Figure \ref{fig:figure6}. The LLM-generated analyses demonstrated high accuracy and safety, as confirmed by the clinicians. However, the study also identified occasional errors in the clinical conclusions produced by the LLMs, which could potentially result in inappropriate treatment decisions~\citep{healey2025case}. While the findings underscore the promise of AI-assisted CGM pattern analysis in enhancing clinical care and clinician efficiency, they also highlight the critical need for further research to refine LLM prompts and integrate human feedback into model training, ensuring greater reliability and clinical applicability~\citep{mallinar2025scalable, yu2023leveraging}.

\section{COMMERCIAL AI-ENHANCED CGM SYSTEMS}
Stelo, the first over-the-counter glucose biosensor cleared by the United States Food and Drug Administration, uses generative AI-enabled technology to produce weekly narrative insights in contextually relevant text. The Stelo app provides personalized tips, recommendations, and education related to diet, exercise, and sleep based on not only glucose data but also meal logs and other wearable data~\citep{Dexcom:2024}. A commercial AI-powered CGM system has been also developed by Roche Diabetes Care. This real-time CGM Diabetes Tracker provides actionable alerts by incorporating AI algorithms to predict glucose highs and lows as well as inform the user of their risk of developing hypoglycemia overnight with their Accu-Chek SmartGuide and SmartGuide Predict app~\citep{glatzer2024clinical}. The app is powered by three machine learning models, including a 120-minute glucose forecast, a 30-minute low glucose detection, and a nighttime low glucose prediction for bedtime interventions~\citep{herrero2024enhancing}. This product is available in some European countries but not in the United States. Other CGMs employ AI for automation and prediction, but do not provide user-facing, generative AI-driven insights at this time.

\section{CGM FOUNDATION MODELS}

A foundation model of CGM is a special type of AI-powered pattern analysis that integrates data, physiological dynamics, and contextual factors to predict glucose patterns, inform clinical decisions, and personalize diabetes management. This type of model is based on deep learning transformer architectures, and it is pretrained on vast amounts of CGM data using self-supervised learning ~\citep{lu2025pretrained}. A foundation model uses a neural network which contains transformers that process data by converting elements (e.g., words or pixels) into numerical representations called tokens. The tokens are then processed in parallel, rather than one at a time, for faster performance~\citep{sergazinov2023gluformer}.
A foundation model of CGM structures for pattern analysis refers to a large, pretrained ML model that learns generalizable representations from CGM data for diverse predictive and analytic tasks~\citep{pyzer2025foundation}. Foundation models capture the underlying dynamics and patterns in the data and then allow building applications, such as diagnosis or risk assessment. Gluformer is an example of a foundation model of CGM data structures. This model can generate CGM data based on dietary intake data, simulate outcomes of dietary interventions, and predict individual responses to specific foods~\citep{lutsker2024glucose}. However, this model is trained primarily on data in the non-diabetic healthy state and is therefore useful only for detecting deviations from a healthy state, i.e., it can detect an increased risk of diabetes.

\section{ML TO PREDICT CLINICAL OUTCOMES FROM CGM}
The Glucose Levels Across Maternity (GLAM) study was an observational, non-interventional study designed to collect CGM data during pregnancy and relate CGM-measured glucose levels with risks of gestational diabetes and preselected perinatal complications~\citep{li2024continuous}. Medication adherence has been determined to be feasible through ML-based pattern analysis of simulated CGM ~\citep{thyde2021machine}. Subtyping patients with T2DM using CGM data may help identify high-risk patients for microvascular complications~\citep{jian2021machine}, including diabetic retinopathy~\citep{shao2024generalization, montaser2024prediction} and albuminuria~\citep{shao2024generalization}. However, these new approaches are at early stages of implementation and require further studies to determine their feasibility and acceptability for use by professionals and people with diabetes. In addition, the potential benefits related to diabetes prevention and reducing the risk of the serious complications associated with diabetes remain to be determined.

\section{CONCLUSIONS}
As Functional Data Analysis-, AI- and ML-enabled applications for CGM for pattern analysis become more available and more powerful, we expect to see greater insights into the user's metabolism and behavior, which will assist clinicians to make more targeted treatment decisions~\citep{sure2025enhancing}. For example, special consideration must be given to populations with irregular circadian rhythms (e.g., shift workers), polypharmacy, or comorbidities that alter metabolic rhythms. These "edge-case" users may not be well-represented in training datasets, and their glucose patterns may challenge standard algorithms. 

As primary care physicians now provide the majority of diabetes care, they are likely to be significant beneficiaries of tools that can translate CGM pattern insights into natural language interpretations. Similarly, and from the perspective of people living with diabetes, personalized insights from more detailed explanations of CGM profiles are very likely to result in improvements in their self-management. Furthermore, for all clinicians, the burden of identifying complex patterns and selecting treatments based on these patterns could be reduced by the same Pattern Analysis 2.0 tools by adding decision support. Researchers will now need to generate practical clinical advice and solutions from their powerful Functional Data Analysis, AI- and ML tools to enable clinicians to move from traditional CGM Pattern Analysis 1.0 to what we are calling CGM Data Analysis 2.0. For example, LLMs could also convert pattern detection into natural language descriptions of recommended treatments to be reviewed and potentially accepted by clinicians as text for clinical documents, which could significantly reduce their workload. Clinicians will soon routinely receive new non-traditional forms of CGM analyses because traditional metrics (CGM Pattern Analysis 1.0) will gradually be replaced by Functional Data Analysis-, AI-, and ML-based reports. These emerging methods for analysis of CGM patterns (CGM Pattern Analysis 2.0) will identify patterns, define the quality of glycemia, and enable truly personalized treatments.

\pagebreak

\section*{DISCLOSURES:}

D.C.K. is a consultant for Afon, Embecta, GlucoTrack, Lifecare, Novo, SynchNeuro, and Thirdwayv.
R.M.B. has received research support, has acted as a consultant, or has been on the scientific advisory board for Abbott Diabetes Care, Ascensia, CeQur, DexCom, Eli Lilly, Embecta, Hygieia, Insulet, Medscape, Medtronic, Novo Nordisk, Onduo, Roche Diabetes Care, Tandem Diabetes Care, Sanofi, United Healthcare, Vertex Pharmaceuticals, and Zealand Pharma.
E.C. has served on the scientific advisory board of Novo Nordisk, Eli Lilly, MannKind, Arecor, Portal Insulin, Provention Bio, Tandem, Sanofi, and Ypsomed.
M.A.C. receives research support from Dexcom and Abbott Diabetes Care and consulting fees from Glooko as Chief Medical Officer
D.E. is a co-founder and shareholder of OneTwo Analytics AB, Sweden.
J.E. receives federal funding from FDA, NIMHD, and NCATS and is a consultant for Sanofi.
D.K.’s institution has received research support from Abbott Diabetes Care.
B.K. reports receiving research support from Dexcom, Inc and Tandem Diabetes Care handled by the University of Virginia; and patent royalties from Dexcom, Inc handled by the University of Virginia’s Licensing and Ventures Group.
D.M.M has had research support from the NIH, NSF, Breakthrough T1D, and the Helmsley Charitable Trust and his institution has had research support from Dexcom. Dr Maahs has consulted for Abbott, Sanofi, Eli Lilly, Medtronic, Biospex, Kriya, and Enable Biosciences.
J.K.M is a member of advisory boards of Abbott Diabetes Care, Becton-Dickinson, Biomea Fusion, DexCom, Eli Lilly, Embecta, Medtronic, myLife, Novo Nordisk A/S, Pharmasens, Roche Diabetes Care, Sanofi-Aventis, Tandem, and Viatris and received speaker honoraria from A. Menarini Diagnostics, Abbott Diabetes Care, DexCom, Eli Lilly, Medtrust, MSD, Novo Nordisk A/S, Roche Diabetes Care, Sanofi, Viatris, and Ypsomed. She is a shareholder of decide Clinical Software GmbH and elyte Diagnostics and serves as CMO of elyte Diagnostics.
N.M. has nothing to disclose.
A.A.M. is currently an employee of Google. 
S.N.S has nothing to disclose.
B.S. has nothing to disclose.
M.P.S. is a co-founder and a member of the scientific advisory board of Personalis, Qbio, January AI, SensOmics, Protos and Mirvie. He is on the scientific advisory board of Danaher, GenapSys and Jupiter. 
G.U. has nothing to disclose.
A.T.A has nothing to disclose.
C.N.H. has nothing to disclose.
E.H. is supported by T32HD040128 from the NICHD/NIH

\pagebreak
\bibliography{cgm}

\renewcommand{\thesubsection}{Supplementary Table S\arabic{subsection}}
\setcounter{subsection}{0} 


\end{document}